\documentclass[conference]{IEEEtran}
\IEEEoverridecommandlockouts
\usepackage{cite}
\usepackage{amsmath,amssymb,amsfonts}
\usepackage{algorithmic}
\usepackage{graphicx}
\usepackage{multirow}
\usepackage{textcomp}
\usepackage{xcolor}
\usepackage{stfloats}
\usepackage{booktabs}

\def\BibTeX{{\rm B\kern-.05em{\sc i\kern-.025em b}\kern-.08em
    T\kern-.1667em\lower.7ex\hbox{E}\kern-.125emX}}

\usepackage{tikz}
\usepackage[hyperfootnotes=false]{hyperref} 

\newcommand\copyrighttext{%
  \footnotesize \textcopyright 2021 IEEE. Personal use of this material is permitted.
  Permission from IEEE must be obtained for all other uses, in any current or future
  media, including reprinting/republishing this material for advertising or promotional
  purposes, creating new collective works, for resale or redistribution to servers or
  lists, or reuse of any copyrighted component of this work in other works.
  DOI: \href{https://doi.org/10.1109/Cluster48925.2021.00052}{https://doi.org/10.1109/Cluster48925.2021.00052}}
\newcommand\copyrightnotice{%
\begin{tikzpicture}[remember picture,overlay]
\node[anchor=south,yshift=10pt] at (current page.south) {\fbox{\parbox{\dimexpr\textwidth-\fboxsep-\fboxrule\relax}{\copyrighttext}}};
\end{tikzpicture}%
}

\newcommand*\concat{\mathbin\big\Vert}
\DeclareMathAlphabet{\altmathcal}{OMS}{cmsy}{m}{n}

\begin{document}

\title{Bellamy: Reusing Performance Models for Distributed Dataflow Jobs Across Contexts}

\author{
\IEEEauthorblockN{Dominik Scheinert, Lauritz Thamsen, Houkun Zhu, Jonathan Will,\\ Alexander Acker, Thorsten Wittkopp, and Odej Kao}
\IEEEauthorblockA{Technische Universit{\"a}t Berlin, Germany, \{firstname.lastname\}@tu-berlin.de}
}

\maketitle
\copyrightnotice

\begin{abstract}
Distributed dataflow systems enable the use of clusters for scalable data analytics. 
However, selecting appropriate cluster resources for a processing job is often not straightforward.
Performance models trained on historical executions of a concrete job are helpful in such situations, yet they are usually bound to a specific job execution context (e.g. node type, software versions, job parameters) due to the few considered input parameters. Even in case of slight context changes, such supportive models need to be retrained and cannot benefit from historical execution data from related contexts.

This paper presents \emph{Bellamy}, a novel modeling approach that combines scale-outs, dataset sizes, and runtimes with additional descriptive properties of a dataflow job.
It is thereby able to capture the context of a job execution. 
Moreover, Bellamy is realizing a two-step modeling approach. 
First, a general model is trained on all the available data for a specific scalable analytics algorithm, hereby incorporating data from different contexts. 
Subsequently, the general model is optimized for the specific situation at hand, based on the available data for the concrete context.
We evaluate our approach on two publicly available datasets consisting of execution data from various dataflow jobs carried out in different environments, showing that Bellamy outperforms state-of-the-art methods.
\end{abstract}


\section{Introduction}
\label{sec:introduction}

Distributed dataflow systems like MapReduce~\cite{Dean2004}, Spark~\cite{Zaharia2010} and Flink~\cite{Carbone2015} allow their users to develop scalable data-parallel programs in a simplified manner, as the parallelism, distribution, and fault tolerance are handled by the respective system. Thereby, the analysis of large volumes of data happens using clusters of computing resources. These resources are commonly managed by resource management systems like YARN~\cite{Vavilapalli2013}, Mesos~\cite{Hindman2011} or Kubernetes\footnote{https://kubernetes.io/}. 

However, the selection of resources and configuration of clusters is often challenging~\cite{Verma2011,Lama2012,Rajan2016}. Even frequent users or experts do not always fully understand system and workload dynamics and thus have difficulties selecting appropriate resources~\cite{Rajan2016,Lama2012}. Meanwhile, there is a growing number of scientists from domains other than computer science who have to analyze large amounts of data every now and then~\cite{Bux2013,Deelman2019}. In light of the increased usage of cloud resources, users can furthermore easily get overwhelmed by the number of possible configurations (e.g. VM types in public clouds). Time and cost budgets are often constrained, which makes it hard to directly find a fitting configuration for the processing job at hand. If processing jobs are accompanied by certain runtime targets, it is typically also required to meet them without spending too much time on finding a suitable resource configuration.

These problems have been addressed following various approaches. Some methods are designed for specific processing frameworks~\cite{Verma2011a,Ferguson2012,AlSayeh2019}, others conduct an iterative profiling strategy~\cite{Alipourfard2017,Hsu2018,Hsu2018a,Hsu2018b}, and a third line of work builds runtime models for evaluating possible configurations. While some works of the third category are based on dedicated profiling runs on a reduced dataset~\cite{Verma2011,Venkataraman2016,Shah2019}, others also incorporate historical runtime data for improved prediction capabilities~\cite{Thamsen2016,Thamsen2017,Verbitskiy2018,Will2021}. Overall, many methods either require a certain amount of historical data, which is not always available, or rely on profiling, which is not always feasible due to budget constraints.

In this work, we  approach the problem of limited training data when building performance models based on historical executions by consideration of cross-context data, i.e. data that originates from executing a job in similar execution contexts. In contrast to the state of the art, which at most considers scale-out information and dataset sizes~\cite{Venkataraman2016,Thamsen2016} and thus only a single context, our novel modeling approach for runtime prediction called \emph{Bellamy} allows for incorporating runtime data from various contexts using additional descriptive properties of a job execution. Such a model is thus reusable across contexts and would therefore work well with scalable data processing in a public cloud~\cite{Will2020}, where in many cases users utilize the same hardware types or algorithm implementations and would hence benefit from sharing information about their job execution.

\emph{Contributions.} The contributions of this paper are:

\begin{itemize}
    \item A novel modeling approach for runtime prediction that incorporates scale-out information as well as other job and resource characteristics for improved prediction capabilities. Using data from various contexts enables the better approximation of an algorithm's scale-out behavior in a specific context. 
    \item An evaluation of our approach to runtime prediction on two publicly available datasets consisting of experimental data from dataflow job executions in different environments. We investigate interpolation and extrapolation capabilities as well as the time required to fit our model.
    \item A prototypical and open source implementation of our approach\footnote{https://github.com/dos-group/bellamy-runtime-prediction}. We provide examples on how to use a trained model for choosing suitable resources.
\end{itemize}

\emph{Outline}. The remainder of the paper is structured as follows.
\autoref{sec:relatedWork} discusses the related work.
\autoref{sec:approach} describes our modeling approach and discusses its advantages. 
\autoref{sec:evaluation} presents the results of our comprehensive evaluation. 
\autoref{sec:conclusion} concludes the paper and gives an outlook towards future work.

\begin{figure*}[bp]
    \centering
    \includegraphics[width=\textwidth]{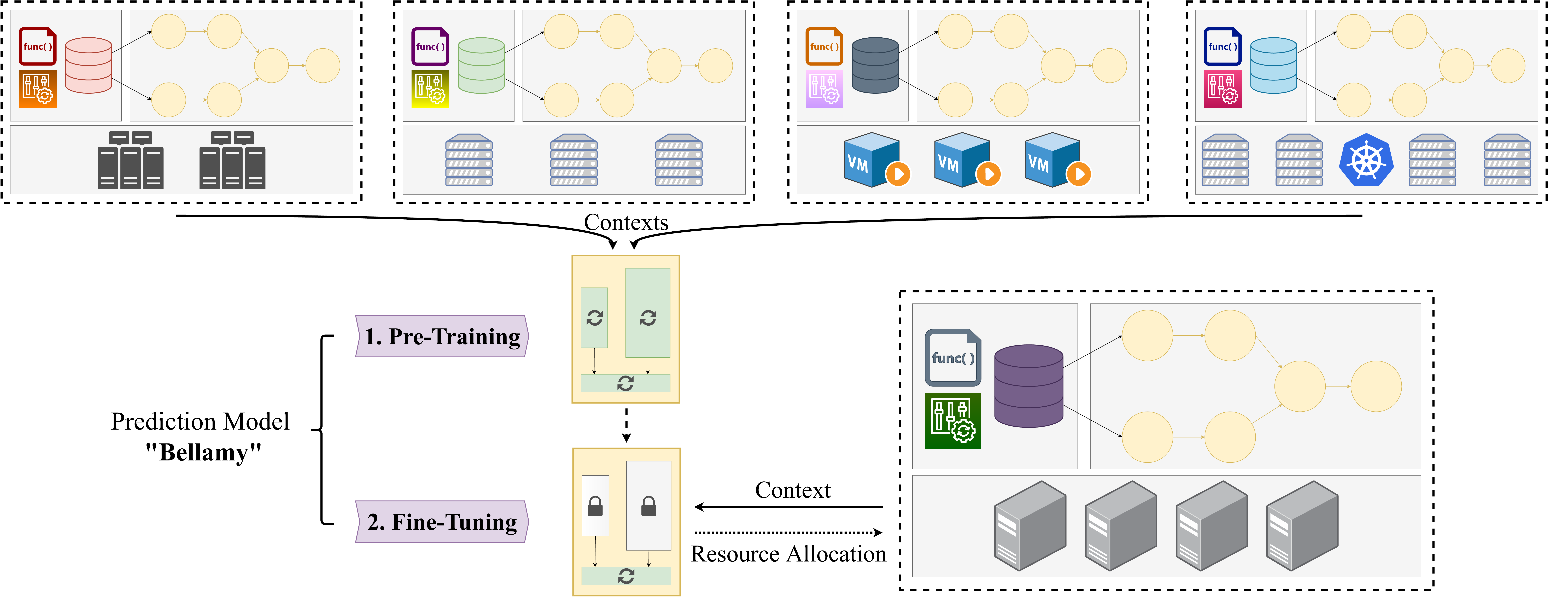}
    \caption{Bellamy learns a job's scale-out behavior model using data from diverse job execution contexts, then optimizes the model for a specific context at hand. In the process, the need for additional profiling runs can be reduced.}
    \label{fig:approach_idea_overview}
\end{figure*}

\section{Related Work}
\label{sec:relatedWork}

Many existing works address runtime prediction for distributed dataflow jobs. They can be categorized into white-box models and black-box models.

\paragraph{White-box models.} These approaches investigate a specific dataflow framework or a class of algorithms, and use white-box models to estimate the runtime.

For example, Apache Spark's multi-stage execution structure is utilized in~\cite{Wang2015} to predict performance. Runtime information from sample runs are collected first and then used to learn the stages behavior to predict job runtimes.

PREDIcT~\cite{Popescu2013} is an approach with focus on predicting the runtime of iterative algorithms. By using sample runs, it captures key information about the convergence pattern and input features in each iteration. Afterwards, it uses those characteristics to predict the runtime of iterative algorithms.

Doppio~\cite{Zhou2018} employs Spark's underlying structure to make predictions. It analyzes the relation between I/O access and computation to build its model, and can be applied on both iterative and shuffle-heavy Spark applications.

Another method strictly designed for Spark is OptEx~\cite{Sidhanta2016}, which employs an analytical modelling approach and incorporates information about the cluster size, number of iterations, the input dataset size, and certain model parameters. 

Some approaches possess characteristics of both classes. A gray-box method is proposed in~\cite{AlSayeh2019}, where a white-box model is used to predict the input RDD sizes of stages under consideration of spark application parameters, while a black-box model utilizes those predicted RDD sizes to predict the runtime of tasks.

Contrary to those models, our approach 
is not specific to a single framework or algorithm, as it is devised as black-box approach.

\paragraph{Black-box models.} Black-box models learn the pattern of dataflow jobs independently of specific frameworks or algorithms. They model the runtime of a job based on training data from dedicated sample runs or preexisting historical runs.

For instance, Ernest~\cite{Venkataraman2016} builds a parametric model, which is trained on a fraction of the real dataset. 
In addition, Ernest uses optimal experiment design to minimize the overhead of training data collection during initial profiling.

Our own previous work Bell~\cite{Thamsen2016} combines a non-parametric model with a parametric model based on Ernest. It trains two models from previous runs, and automatically chooses a suitable model for predictions.

With CherryPick~\cite{Alipourfard2017}, the authors present an approach that selects near-optimal cloud configurations with high accuracy and low overhead. This is achieved by accelerating the process of profiling using Bayesian Optimization, until a good enough solution is found.

Micky~\cite{Hsu2018} improves modeling efficiency with a collective-optimizer, which profiles several workloads simultaneously. To balance the exploration and exploitation, it reformulates the original problem as a multi-arm bandits problem.

Another approach is CoBell~\cite{Verbitskiy2018}, which considers the case of co-located and interfering workloads, and thus trains separate models for different job combinations and considers the interference durations of jobs for the actual runtime prediction.

Tuneful~\cite{Fekry2020} is a recent online configuration-tuning approach which requires no previous training. It utilizes incremental sensitivity analysis and Bayesian optimization to find near optimal configurations.

These approaches can work on general algorithms and frameworks. However, they use few context information as input. We consider other parameters to be useful for runtime prediction too, like node type and job parameters. As a result, our model can adapt to small context changes as it incorporates an understanding of the execution context. This is in contrast to existing methods, which mostly focus on scale-out and dataset information only.

\section{Approach}
\label{sec:approach}
This section presents the main ideas of our approach Bellamy and how it can be used to select appropriate resources according to user-defined runtime targets. We devise a black-box approach in order to potentially apply our solution to multiple distributed dataflow systems.

\subsection{Overview}
\label{sec:approach_overview}

\begin{figure}
    \centering
    \includegraphics[width=\columnwidth]{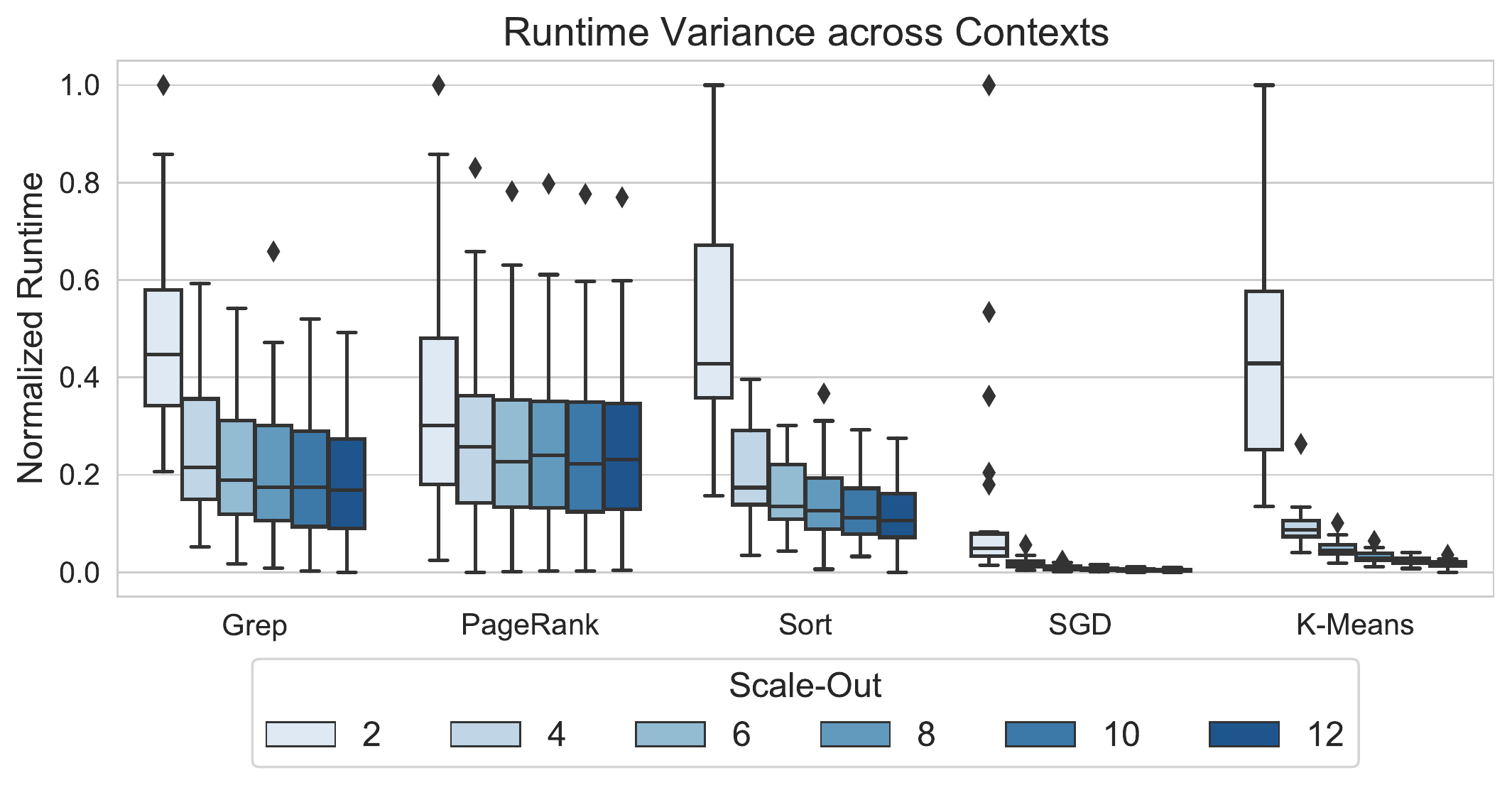}
    \caption{Exemplary illustration of normalized job runtimes in light of different contexts. The reported values originate from our utilized C3O-Datasets, and underline the difficulties of estimating scale-out behaviors of certain jobs.}
    \label{fig:approach_local_runtime_variance}
\end{figure}

Whenever a dataflow job is submitted to a distributed dataflow system, its execution takes place in a specific context. 
As illustrated in~\autoref{fig:approach_local_runtime_variance}, various factors influence the performance of a dataflow job and thus define the context, e.g. the characteristics of the input data, the chosen resources and infrastructure, or implementation details of the respective systems. 
However, it can be observed that many processing algorithms exhibit a similar scale-out behavior, even across contexts~\cite{Will2020}. 
In order to robustly estimate the scale-out behavior of a processing algorithm and predict the runtime of a corresponding concrete dataflow job, we propose to additionally incorporate descriptive properties of the execution context. 
This effectively allows us to potentially learn the scale-out behavior across multiple contexts, as depicted in~\autoref{fig:approach_idea_overview}.

Given a dataflow job, its execution is not only characterized by the horizontal scale-out represented in form of a scalar value $x\in \mathbb{N}$, but also by potentially $p^{(1)}, p^{(2)},\ldots,p^{(n-1)}, p^{(n)}$ numerical or textual descriptive properties of the job execution context. 
Our approach explicitly incorporates the latter in order to utilize data from various contexts, but effectively distinguish them.
We design Bellamy as a neural network architecture which allows for pre-training on a corpus of similar historical execution data, preserving the model state appropriately, and fine-tuning the model as needed for specific use cases.
The model's objective is to jointly minimize the overall runtime prediction error as well as the reconstruction error of the employed auto-encoder for learning latent property encodings.
In order to  fine-tune  a  model,  we  load  the  corresponding  pre-trained  model,  freeze  most model components, and continue the training for a short period of time.

In the subsequent sections, we will describe the individual components of our approach. Our scale-out modeling is introduced in~\autoref{sec:approach_scaleout_modeling}, followed by our approach for encoding descriptive properties of an execution context in~\autoref{sec:approach_configuration_encoding}. Afterwards, we present in ~\autoref{sec:approach_prediction} how the individually obtained intermediate results are effectively combined for predicting the runtime of a dataflow job executed in a specific context. 

\begin{figure}[b!]
    \centering
    \includegraphics[width=.75\columnwidth]{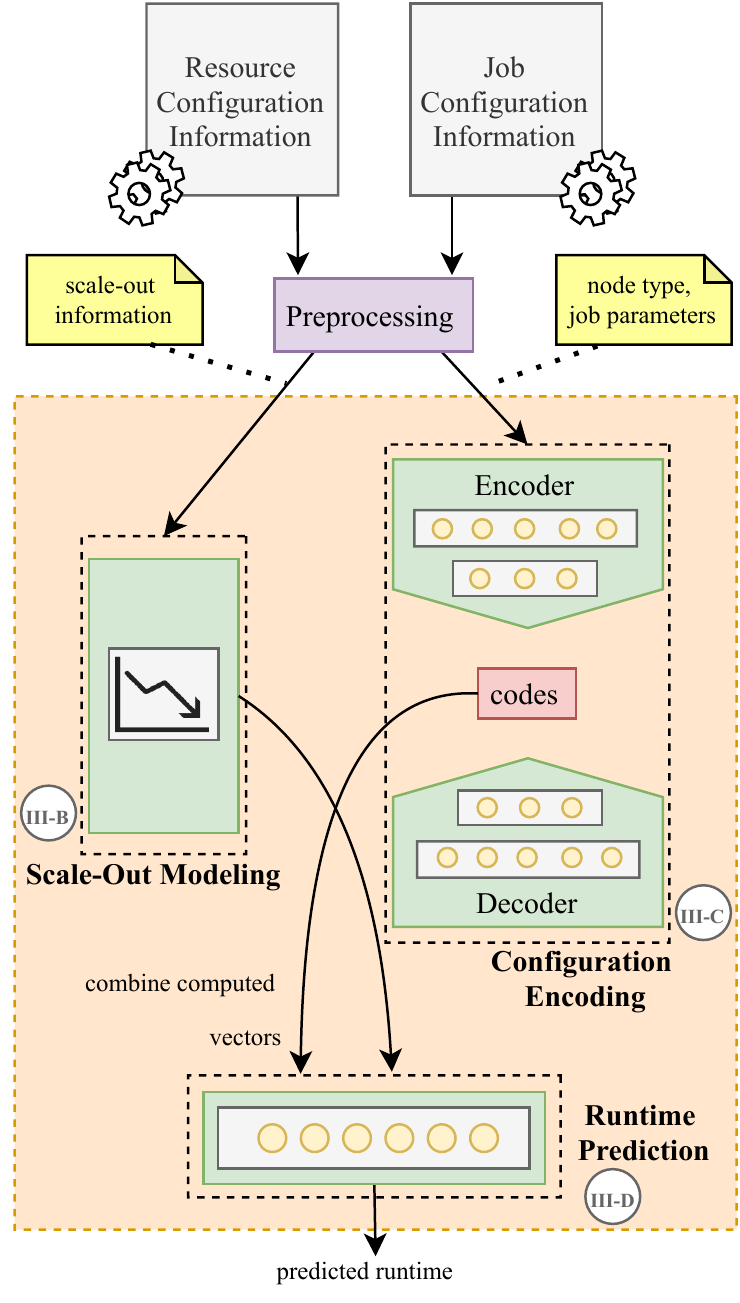}
    \caption{Overview of Bellamy's architecture and general prediction process. The input configurations are obtained from job submission specifications or other sources of information available.}
    \label{fig:approach_model_architecture}
\end{figure}

\subsection{Scale-Out Modeling}
\label{sec:approach_scaleout_modeling}
The parametric model for distributed processing presented with Ernest~\cite{Venkataraman2016} showed to be sufficient for many processing algorithms and their scale-out behavior while maintaining a manageable complexity. It is defined as
\begin{equation}
    f = \Vec{\theta}_1 + \Vec{\theta}_2 \cdot \frac{1}{x} + \Vec{\theta}_3 \cdot \log(x) + \Vec{\theta}_4 \cdot x,
\end{equation}
where each term represents a different aspect of parallel computing and communication, $x$ is the number of machines, and $\Vec{\theta} \in \mathbb{R}^4$ is a vector of weights, which is usually estimated using a non-negative least square (NNLS) solver.
For our scale-out modeling, we borrow from this idea.
Given a scale-out $x\in \mathbb{N}$, we first craft a feature vector $\Vec{x}=[\frac{1}{x}, \log(x), x]^\top$ and use it as input to our transformation function $f: \mathbb{R}^3 \rightarrow \mathbb{R}^F$ to obtain a vector $\Vec{e}\in\mathbb{R}^F$, where $F$ denotes a desired output dimensionality, and $f$ is realized as a two-layer feed-forward neural network.
We choose exactly two layers as this is sufficient to distinguish data that is not linearly separable.

A two-layer feed-forward neural network can be compactly described in a generalized manner with 
\begin{equation}
\Vec{h}_k = \sigma \left( \sum_{j=1}^M \textbf{w}^{(2)}_{kj} \cdot \phi \left( \sum_{i=1}^D \textbf{w}^{(1)}_{ji} \cdot \Vec{x}_i + \Vec{b}^{\,(1)}_j \right) + \Vec{b}^{\,(2)}_k \right),
\end{equation}

where $\sigma$ and $\phi$ denote activation functions, $\Vec{x}\in \mathbb{R}^D$ is the input to the network, $M$ is the output dimension of the first layer (also referred to as hidden dimension of the network), 
$\mathbf{w}^{(1)}\in \mathbb{R}^{M\times D}$ and $\mathbf{w}^{(2)}\in \mathbb{R}^{K\times M}$ are the learnable parameter matrices of the respective layers, 
$\Vec{b}^{\,(1)}\in \mathbb{R}^M$ and $\Vec{b}^{\,(2)}\in \mathbb{R}^K$ are the accompanying optional additive biases, and $\Vec{h}\in \mathbb{R}^K$ represents the output of the network.

For our scale-out modeling component, we utilize such a network with $D=3$ and $K=F$, whereas $M$, $\sigma$ and $\phi$ remain configurable parameters or interchangeable functions. We further refer to the concrete network output as $\Vec{e}$ to be in line with our established definitions.
Eventually, our learnable function $f$ will estimate the scale-out behavior of a certain algorithm based on the initially provided feature vector $\Vec{x}$.

\begin{figure*}
    \centering
    \includegraphics[width=\textwidth]{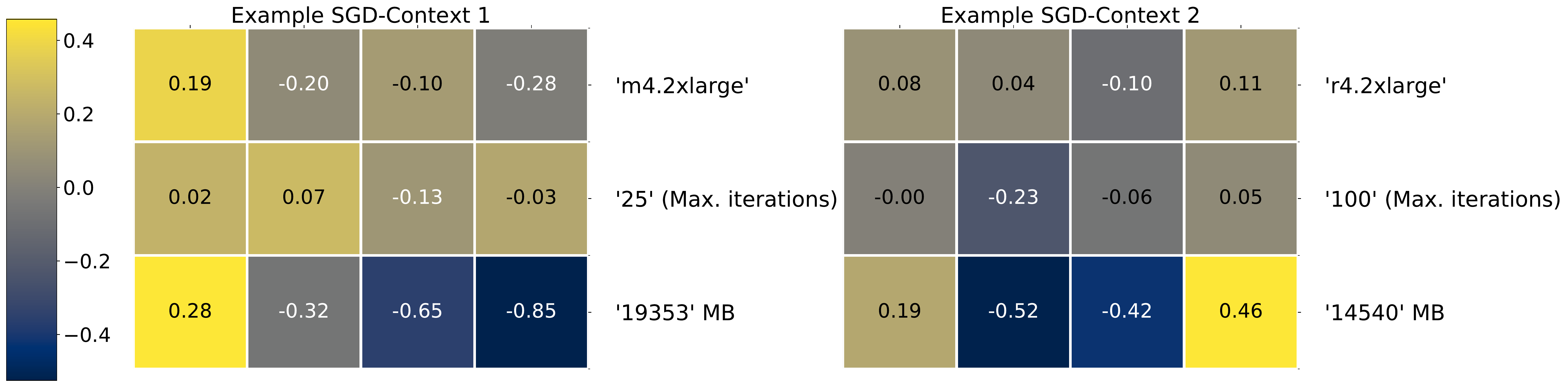}
    \caption{Exemplary visualization of how descriptive properties of two different execution contexts of a SGD job are encoded using our auto-encoder. Each row represents a code. Properties from top to bottom: \emph{node type}, \emph{job parameters} and \emph{dataset size}.}
    \label{fig:approach_examplary_codes}
\end{figure*}

\subsection{Configuration Encoding}
\label{sec:approach_configuration_encoding}
Next to the horizontal scale-out, a job execution is also characterized by a variety of potentially available descriptive properties. Examples are job parameters, the node type, the size of the target dataset, or versions of utilized software. Since certain properties might not be continuously recorded, or are expected to not necessarily add more information (e.g. all executed jobs use the same software version), we distinguish between \textit{essential} and \textit{optional} properties. In case of limited knowledge, each property is regarded as essential.

In order to make use of descriptive properties of a job execution context, we require an efficient, yet robust way of representing these properties. 
In a first step, we transform each property $p^{(i)}$ to a vector of fixed-size length $\Vec{p}^{\,(i)}\in\mathbb{R}^N$, i.e.

\begin{equation}
\Vec{p}^{\,(i)} = [\lambda, \Vec{q}^{\,(i)}_1, \Vec{q}^{\,(i)}_2, \ldots, \Vec{q}^{\,(i)}_{L-1}, \Vec{q}^{\,(i)}_L]^\top,
\end{equation}

where $\Vec{q}^{\,(i)}\in \mathbb{R}^L$ with $L=N-1$ is a vector obtained from an appropriate encoding method as 

\begin{equation}
\Vec{q}^{\,(i)} = \begin{cases}
\text{binarizer}(p^{(i)}) & p^{(i)} \in \mathbb{N}_0 \\
\text{hasher}(p^{(i)}) & \, \text{else}
\end{cases}
\end{equation}

and $\lambda \in \{0,1\}$ is a binary prefix indicating the utilized method.

The \textit{binarizer} method takes a natural number and converts the respective value into its binary representation. 
As a consequence, each property $p^{(i)} \in \mathbb{N}_0$ (e.g. number of CPU cores, memory in MB) can be encoded as long as $p^{(i)} \leq 2^{L}$ holds true.
This saves us the trouble of feature-wise scaling, while allowing for uniquely encoding any number of reasonable size. 

In contrast, the \textit{hasher} method operates on individual textual properties (e.g. job parameters, node type) and follows a different approach. 
First, we strip away all characters that are not part of a user-defined vocabulary. 
Next, we extract different n-grams from the remaining sequence of characters. 
The occurrence of each resulting unique term $t_s$ is then counted and inserted at a specific position in the output vector, such that $\Vec{q}^{\,(i)}_{j} = |t_s|$, where the index $j$ is calculated by the respective hash function that realizes the term to index mapping. 
While collisions for certain computed indices are theoretically possible, it is fairly unlikely that this will happen for all possible indices at once, especially as the textual properties we are working with are comparatively limited in terms of the length of their character sequences.
Lastly, the resulting vector $\Vec{q}^{\,(i)}$ is projected on the euclidean unit sphere such that $\sum_{j=1}^L (\Vec{q}^{\,(i)}_j)^2 = 1$ is ensured. As by this procedure each input property is most likely uniquely encoded, we make the assumption that each input property is predominantly free of errors in the first place (e.g. spelling mistakes), as this would otherwise mean that actually equal inputs are not represented as such. In a practical scenario, this could be ensured by a guided submission tool or automated correction of errors. 

The aforementioned process leads to each property being represented in a way suitable for an algorithm.
However, many of these created vectors can be expected to be sparse, and using them in their raw form would increase the complexity of our model.   
This is why we employ an auto-encoder to obtain dense, low-dimensional representations for each vector.
These so called \emph{codes} are used in conjunction with our scale-out modeling to predict the runtime of a provided dataflow job. 
The auto-encoder is realized using two feed-forward neural networks with two layers each, as defined in~\autoref{sec:approach_scaleout_modeling}.
Given a vector $\Vec{p}^{\,(i)}\in \mathbb{R}^N$, a decoder network function $h:\mathbb{R}^M\rightarrow \mathbb{R}^N$ will try to reconstruct the original vector from the code $\Vec{c}^{\,(i)}\in\mathbb{R}^M$ calculated by the encoder network function $g:\mathbb{R}^N \rightarrow \mathbb{R}^M$, such that $\min\Arrowvert \Vec{p}^{\,(i)} - h(\Vec{c}^{\,(i)}) \Arrowvert_2^2$ and $M \ll N$. The calculated codes can then be used to compactly describe an execution context and to distinguish it from others, as illustrated in~\autoref{fig:approach_examplary_codes}.

\subsection{Runtime Prediction}
\label{sec:approach_prediction}
After obtaining an output from the transformation function $f$ as well as dense property encodings from the encoder network function $g$, we proceed to predict the runtime of the respective dataflow job given its configuration. 
With the encoded context and the enriched scale-out information, we are now able to learn their relationship and to understand the impact on the prospective runtime of the job.
Consider a job execution context with $m$ essential properties, $n$ optional properties, and the corresponding horizontal scale-out, we concatenate the individually computed vectors to a new vector $\Vec{r}\in \mathbb{R}^{F+((m+1)\cdot M)}$ in a capacity-bounded manner, i.e.
\begin{equation}
    \Vec{r} = \Vec{e} \concat \big( \concat_{k=1}^m  \Vec{c}^{\,(k)} \big) \concat \Vec{o}\
\end{equation}
with
\begin{equation}
    \Vec{o}_i = \frac{1}{n}\sum_{j=1}^n \Vec{c}^{\,(j)}_i,
\end{equation}
where $\Vec{e}$ denotes the output vector of the scale-out modeling component,
$(\Vec{c}^{\,(k)})^m_{k=1}$ is a sequence of $m$ codes corresponding to essential properties, and $\Vec{o}$ is the mean vector of $n$ codes corresponding to optional properties.
This way, we enable learning from optional information to some extent, while our model will focus nevertheless on the always available pieces of information.

Eventually, we use a final function $z:\mathbb{R}^{F+((m+1)\cdot M)}\rightarrow \mathbb{R}$ to transform a vector $\Vec{r}$ to a scalar value representing the predicted runtime. Again, we implement $z$ as a two-layer feed-forward neural network. 
During training, our architecture will jointly minimize the overall runtime prediction error as well as the reconstruction error of the employed auto-encoder by accordingly adapting the learnable parameters.
As a result, the function $z$ will be able to distinguish between contexts due to the dense property encodings, understand the effects of individual contexts on the runtime, and nevertheless learn the general scale-out scheme of a certain processing algorithm.

\section{Evaluation}
\label{sec:evaluation}
This section presents  our prototypical implementation, the utilized datasets, and our experiments with accompanying discussion of the results. The implementation and the datasets are provided in our repository\footnote{https://github.com/dos-group/bellamy-runtime-prediction}.

\subsection{Prototype Implementation}
\label{sec:evaluation_prototype_implementation}
Each of our functions, i.e. $f$, $g$, $h$ and $z$, is implemented as a two-layer feed-forward neural network.
Each linear layer is followed by a non-linear activation.
While the last layer of the decoder function $h$ uses a hyperbolic tangent which is in line with the nature of our vectorized properties, we choose the SELU~\cite{Klambauer2017} activation function for all other layers, as it has been shown to not face vanishing and exploding gradient problems while still speeding up training and improving the overall generalization performance of a model. All parameters in our functions are thus initialized using He initialization~\cite{He2015} in accordance with the specific properties of our activation.

\begin{table}[b]
\centering
\caption{Model Configuration and Training}
\begin{tabular}[t]{c||c}
    \toprule
    \multicolumn{2}{c}{\textbf{Configuration}}\\
    \toprule
    \multirow{2}{*}{General} & Hidden-Dim. = 8, Out-Dim. = 1\\
        &  Decoding-Dim. = 40, Encoding-Dim. = 4\\ 
    Batch size & 64\\
    Optimizer & Adam\\
    \midrule
    \multicolumn{2}{c}{\textit{Pre-Training}}\\
    \midrule
    Loss & Huber (Runtime) + MSE (Reconstruction)\\
    Dropout & \{5\%, 10\%, 20\%\}\\
    Learning rate & \{$1e^{-1}, 1e^{-2}, 1e^{-3}$\}\\
    Weight decay & \{$1e^{-2}, 1e^{-3}, 1e^{-4}$\}\\
    \#Epochs & 2500\\
    \midrule
    \multicolumn{2}{c}{\textit{Fine-Tuning}}\\
    \midrule
    Loss & Huber (Runtime)\\
    Dropout & 0\%\\
    Learning rate & cyclical annealing in $(1e^{-2}, 1e^{-3})$\\
    Weight decay & $1e^{-3}$\\
    \#Epochs & max. 2500 \\
    Stopping criterion & MAE $\leq 5$, or no improvement in 1000 epochs\\
    \bottomrule
\end{tabular}
\label{tbl:evaluation_experiment_setup}
\end{table}

The input to $f$ is normalized to the range $(0, 1)$ feature-wise, where the boundaries are determined during training and used throughout inference.
For the initial transformation of descriptive properties into vectors, we choose a vector size of $N=40$ in order to allow for encoding larger numbers while also reducing the collision probability of the utilized hash function. 
Encoding natural numbers is straightforward using the aforementioned binary transformation.
For textual properties, we first utilize a simple case insensitive character-vocabulary with alphanumeric characters and a handful of special symbols. Characters not present in the vocabulary are stripped away. 
We then extract unigrams, bigrams, and trigrams from the cleaned textual properties, and eventually use the \textit{HashingVectorizer} from scikit-learn\footnote{https://scikit-learn.org/0.23/index.html}.

We configure the encoder function $g$ with an input dimension of 40, a hidden dimension of 8, and an output dimension of 4. 
The same applies to the decoder function $h$ but in reverse order. 
Both functions waive additional additive biases, and also utilize an alpha-dropout~\cite{Klambauer2017} mechanism during training between their respective layers to mitigate overfitting. 
Our scale-out function $f$ has by design a fixed input dimension of 3, a hidden dimension of 16, and an output dimension of 8. 
Lastly, $z$ gradually maps to the desired output dimension of 1 by utilizing a hidden dimension of 8.

In our experiments, we obtain a pre-trained model after a hyperparameter search. The search space is depicted in \autoref{tbl:evaluation_experiment_setup}, and we sample 12 configurations from it using Tune~\cite{Liaw2018} with Optuna~\cite{Akiba2019}. More details can be found in the aforementioned repository. 
Whenever we attempt to fine-tune a model, we continue the model training on the respective data samples from a new concrete context. In the process, we first update only parameters of the function $z$, while also allowing to update the parameters of function $f$ after a number of epochs dependent on the amount of data samples.
We keep track of the best model state according to the smallest runtime prediction error and use this model state afterwards for inference. 
We prematurely finish the fine-tuning if the mean absolute error (MAE) of the runtime prediction is smaller or equal a specified value, or the error did not decrease in a defined range. This is further described in \autoref{tbl:evaluation_experiment_setup}.

\subsection{Datasets}
\label{sec:evaluation_datasets}
We utilize datasets originating from distinct environments.

\paragraph{C3O-Datasets} We use the datasets\footnote{https://github.com/dos-group/c3o-experiments} provided with the corresponding paper~\cite{Will2020}, where we conducted 930 unique runtime experiments of distributed dataflow jobs with five different algorithms in a \emph{public cloud environment}, i.e. Amazon EMR which uses Hadoop 3.2.1 and Spark 2.4.4.
For the C3O-datasets, an execution context is uniquely defined by the node type, job parameters, target dataset size, and target dataset characteristics.  
There are 21 unique execution contexts for Sort, 27 for Grep, 30 for each SGD and K-Means, and 47 for PageRank. 
For each context, 6 scale-outs were investigated ranging from 2 to 12 machines with a step size of 2. 
The experiment for each scale-out was repeated 5 times.  

\paragraph{Bell-Datasets} We make use of the datasets\footnote{https://github.com/dos-group/runtime-prediction-experiments} provided with~\cite{Thamsen2016}, where we conducted the corresponding experiments in a \emph{private cluster environment} with Hadoop 2.7.1 and Spark 2.0.0. We select the results of three utilized algorithms (Grep, SGD, PageRank), each executed in a single context. For each context, 15 scale-outs were investigated ranging from 4 to 60 machines with a step size of 4. The experiment for each scale-out was repeated 7 times.

Using these datasets, we select \emph{dataset size}, \emph{dataset characteristics}, \emph{job parameters}, and \emph{node type} as essential input properties, as well as \emph{memory} (in MB), \emph{number of CPU cores}, and \emph{job name} (e.g. SGD) as optional input properties.

\subsection{Experiments}
\label{sec:evaluation_experiments}
The Pre-Training of Bellamy models was conducted on a dedicated machine equipped with a GPU. 
Normal training or fine-tuning of models was conducted using the CPU only. 
Specifications and software versions can be found in \autoref{clusterspecs}. 

\begin{table}[ht]
\centering
\caption{Hardware \& Software Specifications}
    \begin{tabular}[t]{rp{0.68\linewidth}}
        \toprule
        Resource&Details\\
        \midrule
        CPU, vCores & Intel(R) Xeon(R) Silver 4208 CPU @ 2.10GHz, 8\\
        Memory & 45 GB RAM\\
        GPU & 1 x NVIDIA Quadro RTX 5000 (16 GB memory)\\
        Software & PyTorch 1.8.0, PyTorch Ignite 0.4.2\\ 
        & PyTorch Geometric 1.7.0, Ray Tune 1.1.0\\
        & Optuna 2.3.0, scikit-learn 0.23.2\\
        \bottomrule
    \end{tabular}
\label{clusterspecs}
\end{table}

We compare our black-box and model-based approach Bellamy to the most related state of the art methods, namely the parametric model of Ernest~\cite{Venkataraman2016} and our own previous work Bell~\cite{Thamsen2016}.
In the process, we investigate their interpolation and extrapolation capabilities as well as the time required for fitting the respective models.
We are especially interested in the performance of our approach when only a limited number of data samples is available for a concrete context. 
This is motivated due to the fact that each data sample is the result of a job execution, which in turn means that models that require much data are unfavorable as they introduce additional costs when recording an initial set of data samples.
Thus, we evaluated the prediction performance of all models with different numbers of available training data points. 
Given a concrete job execution context, for each model and number of training data points we calculated the respective prediction error using random sub-sampling cross-validation. 
For every fixed amount of training data points, random training points are selected from the dataset such that the scale-outs of the data points are pairwise different. 
To evaluate the interpolation capabilities of all models, we then randomly select a test point such that its scale-out lies in the range of the training points. 
For evaluating the extrapolation capabilities, we randomly select a test point such that its scale-out lies outside of the range of the training points. 
The prediction errors are eventually calculated by comparing the predicted runtimes with the actual runtimes. 

\begin{figure*}
    \centering
    \includegraphics[width=0.85\textwidth]{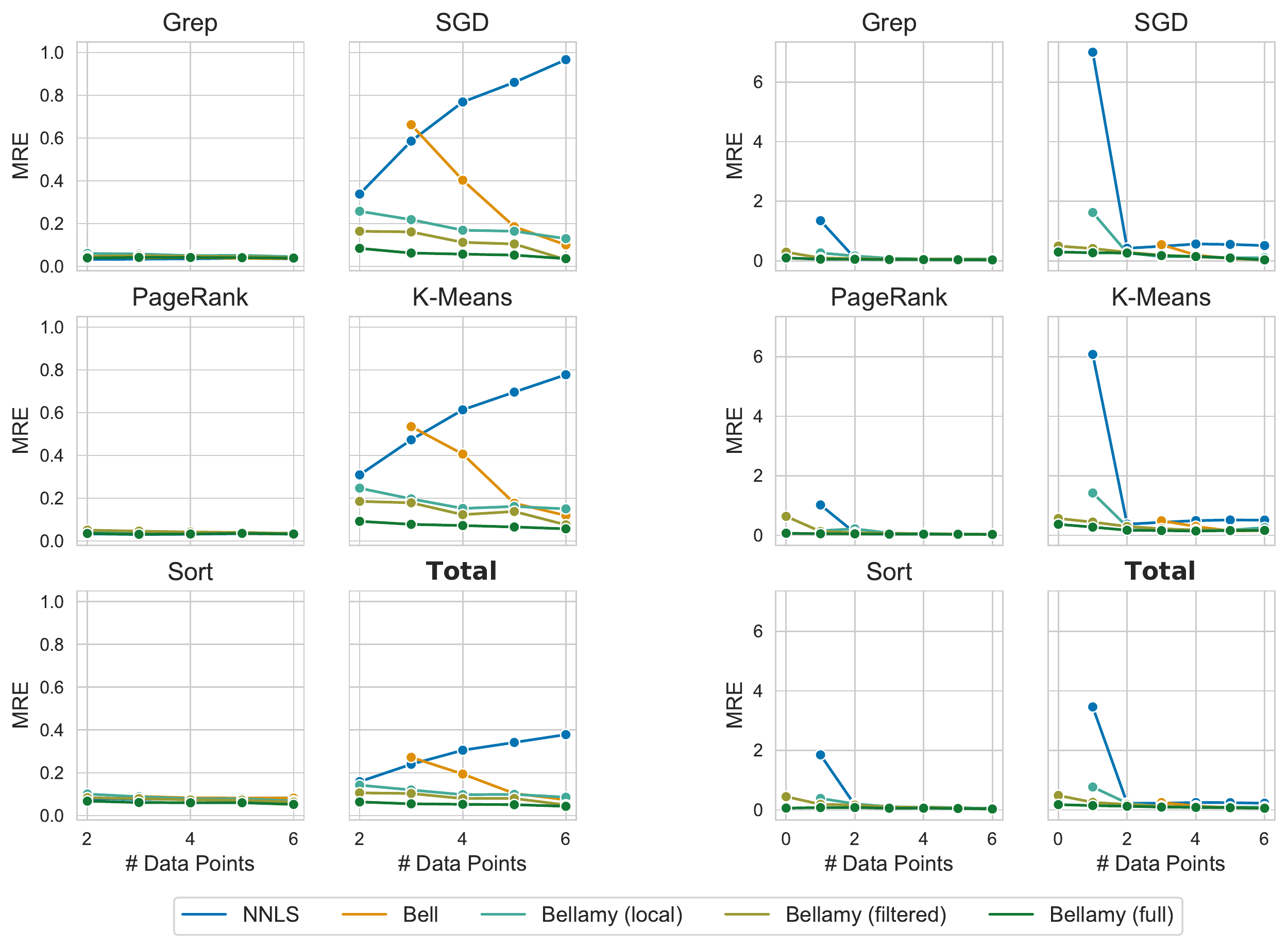}
    \caption{Ad Hoc Cross-Context Learning. \textbf{Left:} Mean relative errors (MRE) on the task of \emph{interpolation} across splits and contexts. On average, pre-trained Bellamy models tend to interpolate better. \textbf{Right:} Mean relative errors (MRE) on the task of \emph{extrapolation} across splits and contexts. Pre-trained Bellamy models overfit less on the provided context.}
    \label{fig:evaluation_local_interpolation_extrapolation}
\end{figure*}

\subsubsection{Ad Hoc Cross-Context Learning}
In this series of experiments, we use the C3O-datasets and investigate the potential of learning from data that originates from different execution contexts.
The aforementioned sub-sampling procedure is repeated as long as we obtain at most 200 unique splits (interpolation test, training, extrapolation test) for each amount of training points.
In order to arrive at a meaningful comparison, we investigate different variants of our approach Bellamy. 
Consider a concrete job in a new and specific context, then we investigate three different pre-training scenarios:
\begin{itemize}
    \item{\textit{local}}: No historical data from different contexts is available and thus no pre-training is possible. Consequently, the auto-encoder is not trained as it bears no advantage.
    \item{\textit{filtered}}: We pre-train our model on historical executions of the same job where the contexts are as different as possible to the one at hand, i.e. we only incorporate data from contexts where the node type, data characteristics, and job parameters do not match and the dataset size is either significantly larger or smaller ($\geq 20\%$). We thus investigate if there is value in learning from historical data that originates from substantially different contexts.
    \item{\textit{full}}: We pre-train our model on all historical executions of the same job in different contexts. This might encompass both similar and distinct contexts.
\end{itemize}
The respective model is eventually fitted / fine-tuned solely on the provided data samples from the new context.
We repeat the described procedure for 7 randomly chosen different contexts for each job, assuring that each node type is present at least once in one of the contexts. Both prediction errors and runtimes for fitting the models are then averaged across the chosen contexts and splits.

\textbf{Interpolation.} The plots on the left-hand side in \autoref{fig:evaluation_local_interpolation_extrapolation} show the mean relative errors (MRE) for the task of interpolation. 
As expected with increasing amounts of training data points
and hence higher density of the dataset, the interpolation capabilities of the non-parametric models surpass the ones of the parametric model. 
It can be seen that pre-training on data from other contexts generally enables the respective Bellamy variants to constantly perform better. 
For algorithms with a non-trivial scale-out behavior (in this example K-Means and SGD), this manifests in significant differences in terms of mean relative errors.
A good prediction performance for small amounts of data points is important, as it leads to less initial profiling and thus saves resources which are often constrained anyway. 
In contrast, all models achieve comparably good results for algorithms (in this example Sort, Grep, PageRank) where the observable scale-out behavior is rather trivial. 
The Bellamy variant without any pre-training is on average inferior to the pre-trained variants.

\begin{figure}
    \centering
    \includegraphics[width=\columnwidth]{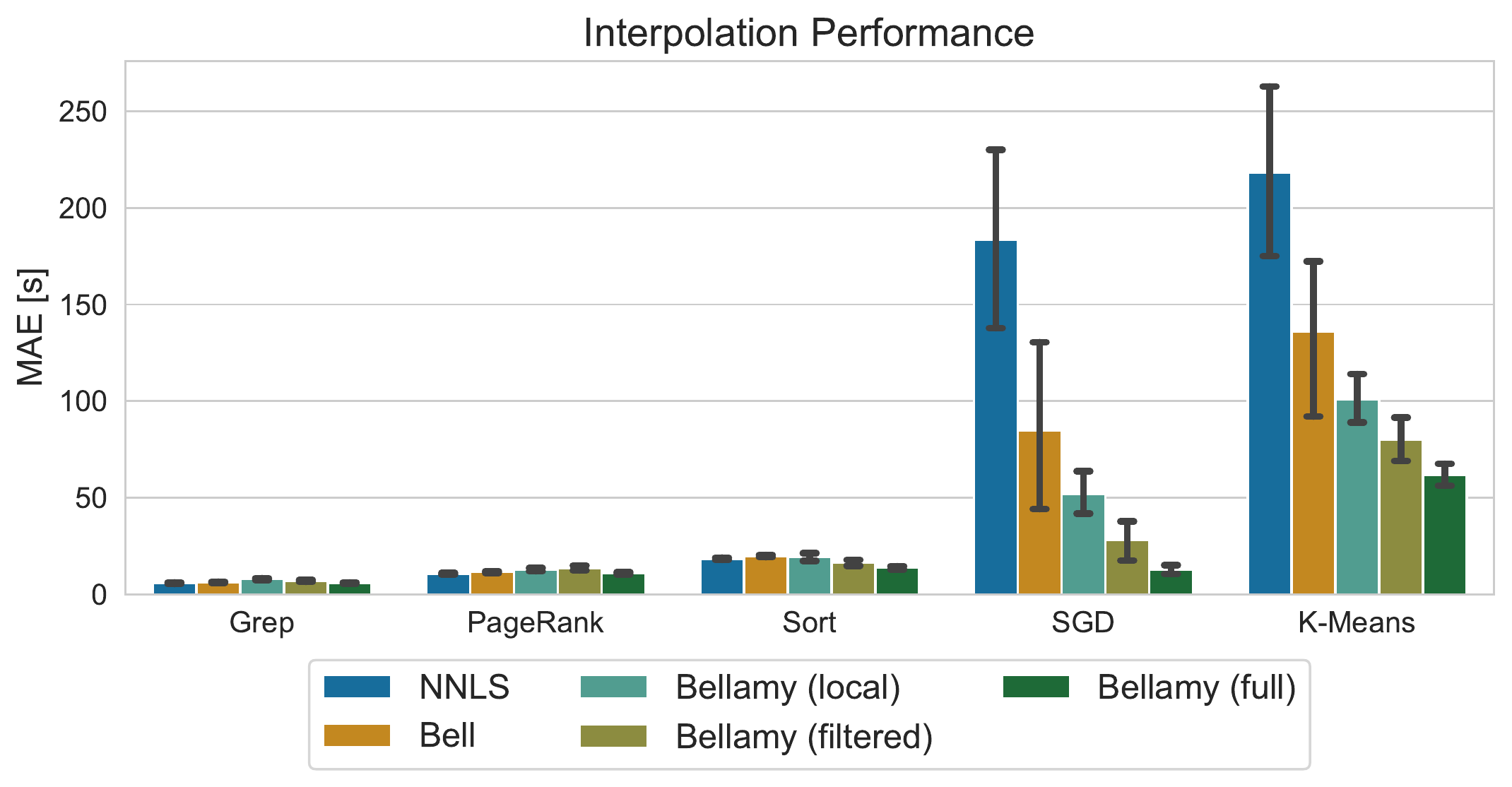}
    \caption{Ad Hoc Cross-Context Learning. Mean absolute errors (MAE) on the task of \emph{interpolation} across splits, contexts, and number of investigated data points. Though all variants are on par or superior to the comparative methods, using pre-trained Bellamy variants leads to stable and improved prediction results.}
    \label{fig:evaluation_local_avg_prediction_error}
\end{figure}

Further, we summarize the models interpolation capabilities by means of the mean absolute errors (MAE). 
\autoref{fig:evaluation_local_avg_prediction_error} shows the results, and in the process highlights the differences in prediction performance, which maximize for algorithms with non-trivial scale-out behavior. 
It can also be observed that our approach Bellamy is more stable across investigated contexts and number of data points. 
While the mean absolute errors in parts already amount to minutes in our experiments, it is self-evident that the errors will increase even further for larger datasets that need to be processed by a dataflow job.  

It is in general highly desirable to utilize a prediction method that not only performs well with small amounts of data points, but also keeps the prediction error manageable. 
Since methods like NNLS or Bell are eventually used for selecting a suitable scale-out that meets certain runtime targets, an inaccurate model can favor the selection of not ideal resources, which in turn can introduce unnecessary costs.
We find that our approach Bellamy obeys these requirements.

\textbf{Extrapolation.} The plots on the right-hand side in \autoref{fig:evaluation_local_interpolation_extrapolation} report the extrapolation results. 
It can be observed that our baselines require a certain amount of data points for adequate results. 
For instance, using NNLS with just one data point is by design unreasonable, whereas Bell requires at least three data points due to an internally used cross-validation. 
In contrast, a pre-trained Bellamy model can be directly applied in a new context without any seen data points, as illustrated in the plot. 
Although it can be seen that fine-tuning on an increasing number of data samples helps to reduce the extrapolation error, the latter is already manageable in many cases without any fine-tuning at all.

These findings are again especially useful in the context of limited data points or constrained resources. 
Being enabled to directly apply a pre-trained model without any initial profiling, or to achieve good enough extrapolation results for small amounts of data, is of advantage in such use cases.

\begin{figure*}
    \centering
    \includegraphics[width=\textwidth]{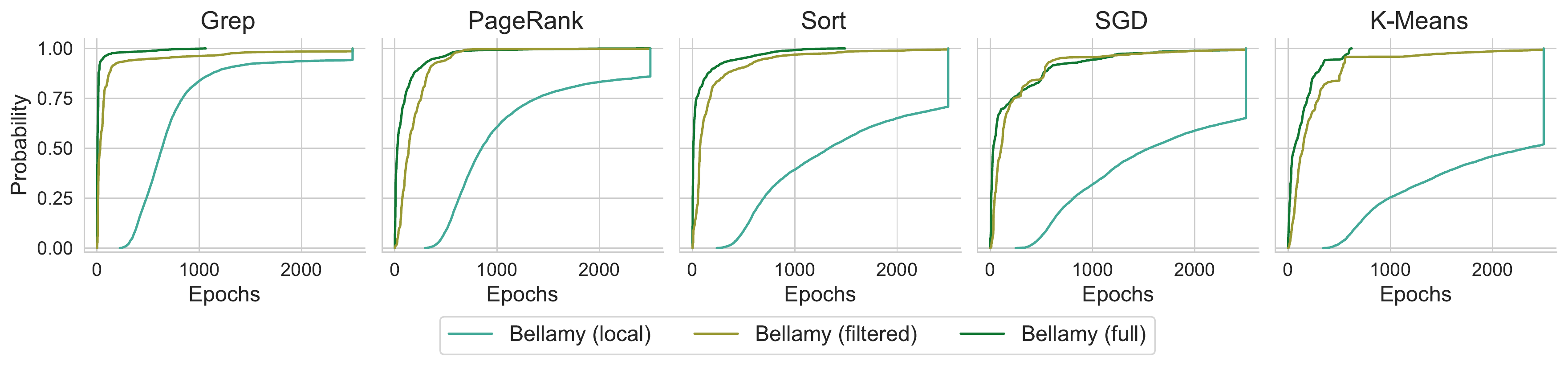}
    \caption{Ad Hoc Cross-Context Learning. Empirical cumulative distribution function (eCDF) of trained epochs for each algorithm and variant of Bellamy. Pre-trained Bellamy models converge faster and thus terminate the fine-tuning earlier, which significantly impacts on the required overall training time.}
    \label{fig:evaluation_local_finetuning_runtime}
\end{figure*}

\textbf{Training time.} In our experiments, fitting both NNLS and Bell on a handful of data points took at most a few milliseconds. In contrast, we observed an average time to fit of $7.37$s for the \textit{local}, $0.99$s for the \textit{filtered}, and $0.55$s for the \textit{full} variant of Bellamy. 
These average runtimes also include the preparation of the respective training pipelines and, if the case, loading a pre-trained model from disk. 
For each variant of Bellamy, we found a considerable amount of outliers with regards to the runtime, which are partially a result of our chosen grace period before termination, and the fact that we calculate the average training time over all experiments and number of data points. 
Consequently, the time varies dependent on the number of data points.
\autoref{fig:evaluation_local_finetuning_runtime} allows for more insights as it illustrates the empirical cumulative distribution function (eCDF) of trained epochs for each algorithm and variant of Bellamy. 
Not surprisingly, it can be seen that the pre-trained variants are converging and hence terminating significantly faster than the \textit{local} variant. A large proportion of experiments finishes within few hundred epochs, which is in line with the aforementioned mean runtimes. In contrast, the amount of epochs required without any pre-training is often volatile. This is underlined by many experiments not finishing prematurely at all, as indicated by the last \textit{jump} of the local Bellamy variant. Moreover, it can be observed that all variants require more training when the scale-out behavior inherent to the experiments conducted for a certain algorithm is not trivial. This is evidently demonstrated when comparing the eCDF of a model variant horizontally across processing algorithms.

While more time consuming than our baselines, the explored prediction advantages should in most cases outweight the introduced and often negligible training overhead, especially for long running dataflow jobs. 

\subsubsection{Potential of Ad Hoc Cross-Environment Learning}
We use both datasets in this series of experiments and investigate the potential of reusing models that were trained on data from a different environment, which potentially implies a significant context shift. More precisely, we simulate the use case of migrating from a public cloud environment (models trained on data from C3O-datasets) to a private cluster environment (data from Bell-datasets), which implies changes in utilized hardware, software, and infrastructure setup.
For each algorithm present in both datasets, we first obtain a pre-trained Bellamy model using the C3O-datasets, and then proceed to directly reuse it on data associated with the Bell-datasets. 
The aforementioned sub-sampling procedure is repeated as long as we obtain at most 500 unique splits (interpolation test, training, extrapolation test) for each amount of training points.
Furthermore, we investigate different ways of reusing the pre-trained models:
\begin{itemize}
    \item{\textit{partial-unfreeze}}: The parameters of function $z$ are adapted, later on also the parameters of function $f$.
    \item{\textit{full-unfreeze}}: The Parameters of function $f$ and $z$ are both adapted from the start.
    \item{\textit{partial-reset}}: We re-initialize the parameters of function $z$ and fine-tune the model, e.g. in order to overcome a previously found local minimum.
    \item{\textit{full-reset}} Parameters of function $f$ and $z$ are re-initialized, i.e. we allow for deriving a new understanding of the scale-out behavior.
\end{itemize}
In each of the above cases, the parameters of our auto-encoder are not subject to changes.
We also use a \textit{local} Bellamy model for comparison. Apart from that, the rest of our experiment design is similar to the one of the previously described experiment, with the exception of us only having access to a single context for each algorithm due to the nature of the Bell-datasets.

\begin{figure}
    \centering
    \includegraphics[width=\columnwidth]{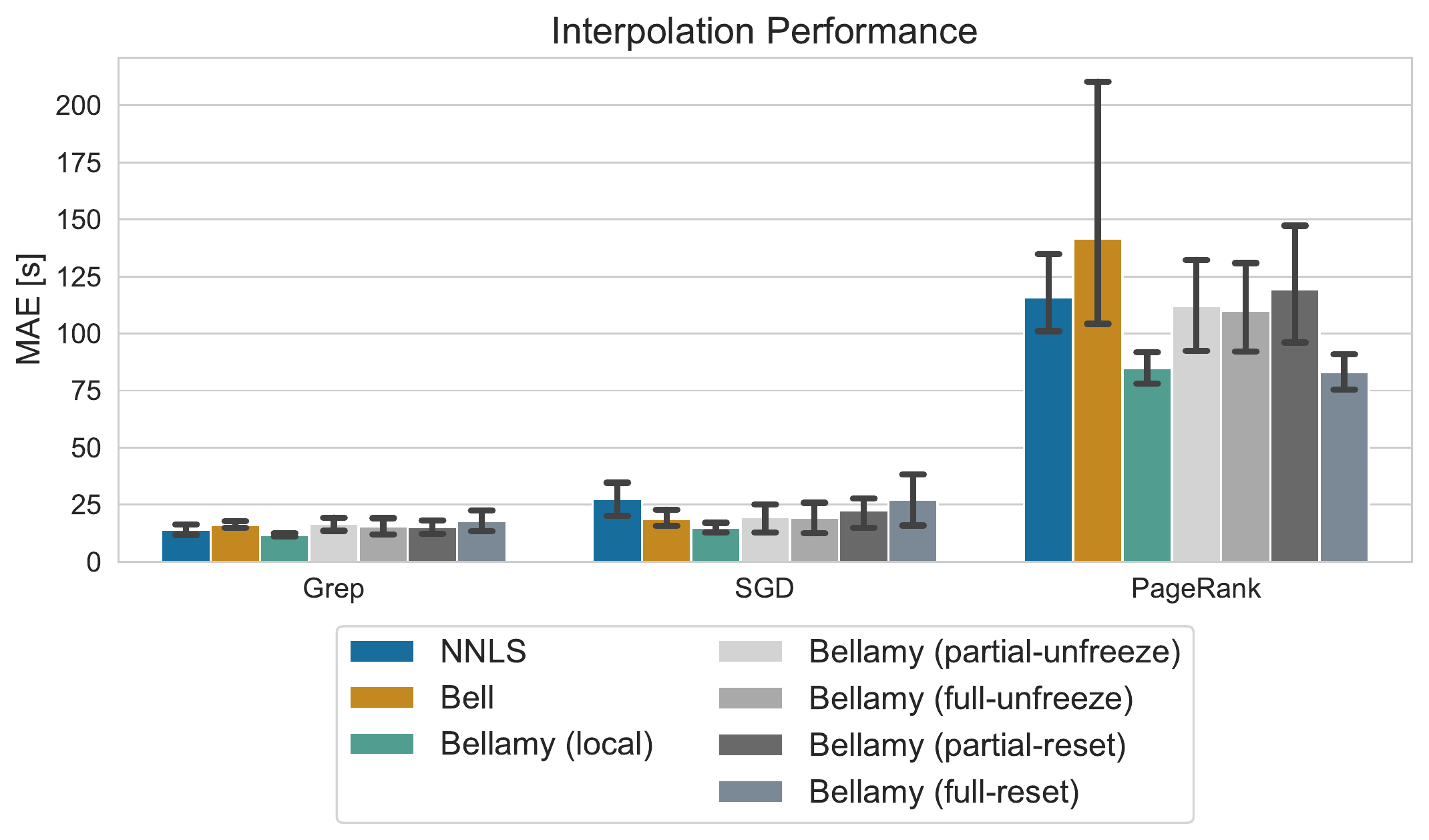}
    \caption{Ad Hoc Cross-Environment Learning. Mean absolute errors (MAE) on the task of \emph{interpolation} across splits and number of investigated data points are reported. Potential of reusing models can be discovered for algorithms with a non-trivial scale-out behavior.}
    \label{fig:evaluation_baremetal_interpolation_tf}
\end{figure}

The interpolation results for the three algorithms (Grep, PageRank, SGD) are summarized in~\autoref{fig:evaluation_baremetal_interpolation_tf}. Similar to the first series of experiments, we find that there are general differences in how good the scale-out behavior of an algorithm can be estimated. For Grep and SGD, all models perform comparably well, with some being slightly more stable than the rest. In contrast, the prediction performance of all models is worse for PageRank, while at the same time revealing significant differences between models. For instance, it can be observed that both the \textit{local} as well as the \textit{full-reset} Bellamy variant exhibit superior performance while also being the most stable. All other investigated Bellamy variants are less stable, and are mostly on par with the parametric model (NNLS). Across all three algorithms, the \textit{local} variant shows on average the best prediction performance. We generally observe that the Bellamy variants that try to make use of the already trained weights experience difficulties.

As for the required training time, we find that all variants based on a pre-trained model exhibit mean runtimes between $2.8$s and $3.8$s, whereas the local variant has a mean runtime of $9.4$s. 
Therefore, if the prediction performance of a pre-trained model is similar to the one of the local model, it is worth considering using the pre-trained model to speed up the training process.

\subsection{Discussion}
\label{sec:evaluation_discussion}
Our main investigation on the C3O-datasets revealed that our approach allows for improved prediction results when incorporating historical data of related contexts, which is especially useful for processing algorithms with a non-trivial scale-out behavior. 
Since the C3O-datasets originate from experiments that emulate job executions from diverse users in the same environment, Bellamy qualifies for being utilized by users with infrequent processing needs, e.g. in a public cloud. 
This way, users can profit from historical data of differently configured job executions. 
A collaborative system for sharing historical execution data across users would favor our approach even more. We also find that good results are achievable with a few data points already, which minimizes the costs for recording an initial dataset of historical executions.

Our second series of experiments investigated the extreme case of ad hoc reusing a model in another environment, i.e. under substantially different conditions which implies a significant context shift. While a pre-trained model does not necessarily lead to superior overall prediction performance, we observe that it can accelerate the training and is therefore a valid option. This bears the potential of benefiting from historical execution data even after situations like infrastructure migration or major software updates.
It is in general advisable to describe the enclosing job execution context of of a dataflow job appropriately when using data from diverse contexts and even environments, such that a Bellamy model can understand the relationship between contexts and corresponding runtimes.

For algorithms with presumably trivial scale-out behavior, we observed that Bellamy models were not always superior to our utilized baselines. 
On the one hand, this is partially a result of our relaxed stopping criterion for the training and the lack of data for proper early stopping. As a consequence, the training might be terminated before an optimal solution was found.
On the other hand, if the scale-out behavior of an algorithm is rather trivial, e.g. when it is presumably linear, our employed baselines are also enabled to provide accurate estimates, while having fewer parameters to train which makes it less likely to find only a near-optimal solution. As a result, the advantage of our approach is more evident for algorithms with presumably non-trivial scale-out behavior.

\section{Conclusion}
\label{sec:conclusion}
This paper presented \emph{Bellamy}, a novel modeling approach for predicting the runtimes of distributed dataflow jobs that allows for incorporating data from different contexts. 
The predicted runtimes can be used to effectively choose a suitable resource configuration for a specific job in a particular execution context.
Bellamy not only uses information about scale-outs and dataset sizes, but also incorporates additional descriptive properties of a job execution context and thus allows to learn models using data from different contexts. 
Despite the consideration of additional descriptive properties, Bellamy is nevertheless a black-box approach, as it does not require detailed statistics or monitoring data, and as a consequence can be used with different resource managers and for different dataflow systems.

We implemented Bellamy as a neural network with multiple task-specific components. 
As shown by our evaluation on publicly available datasets, Bellamy is able to interpolate the scale-out behavior of a dataflow job better than state-of-the-art methods, in the process making use of historical execution data from a variety of contexts. 
The advantage of our approach is especially significant for processing algorithms with non-trivial scale-out behavior. 
We also observed potential when reusing models across vastly different environments.

In the future, we want to investigate possibilities of incorporating dataflow graph information into the prediction process. 
Moreover, since some processing algorithms showed a similar scale-out behavior, we further plan to research ways of building models across algorithms.

\section*{Acknowledgments}
This work has been supported through grants by the German Federal Ministry of Education and Research (BMBF) as BIFOLD (funding mark 01IS18025A).

\bibliographystyle{IEEEtran}
\bibliography{bib}

\end{document}